\documentclass[aps]{revtex4}
\renewcommand\ln{\log}
\newcommand\ba{\begin{eqnarray}}
\newcommand\ea{\end{eqnarray}}

\usepackage{graphicx}
\usepackage{dcolumn}
\usepackage{bm}

\begin{document}


\title{
Weak radiative corrections to the Drell-Yan process\\
for large invariant mass of a dilepton pair}

\author{Vladimir A. Zykunov}
\email{zykunov@sunse.jinr.ru, zykunov@gstu.gomel.by}
\affiliation{
Joint Institute for Nuclear Research, 
141980, 
Dubna, Russia\\
and\\
Gomel State Technical University\mbox{,}
246746, 
Gomel, Belarus\\
}%

\date{\today}

\begin{abstract}
The weak radiative corrections to the Drell-Yan process
above the $Z$-peak have been studied.
The compact asymptotic expression for
the two heavy boson exchange --
one of the significant contributions to the investigated process --
has been  obtained,
the results expand in the powers of the Sudakov electroweak logarithms.
At the quark level we compare the weak radiative corrections to
the total cross section and forward-backward asymmetry with the
existing results and achieve a rather good coincidence
at $\sqrt{s} \gtrsim $ 0.5~TeV.
The numerical analysis has  been performed in the high energy region
corresponding to the future experiments
at the CERN Large Hadron Collider (LHC).
To simulate the detector acceptance we used the standard CMS detector cuts.
It was shown that double Sudakov logarithms of the WW boxes are  
the dominant contributions in hadronic cross section.
The considered radiative corrections are significant
at high dilepton mass $M$ and change the dilepton mass distribution
up to $\sim +3 (-12) \% $
at the LHC energies and $M=1(5)\mbox{TeV}$.
\end{abstract}


\maketitle

\section{Introduction}
Despite the fact that the Standard Model (SM) for more than twenty years
has been considered a consistent and experimentally confirmed theory, 
the search of New Physics (NP) manifestations has still been continued.
As the possible (and, at least, most popular) phenomena of NP, we can name
the supersymmetry, the large extra dimensions \cite{extra-dim} and
extra neutral gauge bosons \cite{extra-bos}.
The light on this  paramount problem of modern physics
can be shed by the coming in the near future
experiments at the collider LHC.

The experimental investigation of the continuum
for the Drell-Yan production of a dilepton pair, i.e.
data on the cross section and  the forward-backward asymmetry of
the reaction
\begin{equation}
pp \rightarrow \gamma, Z \rightarrow l^+l^-X
\label{1}
\end{equation}
at large invariant mass of a dilepton pair (see \cite{cmsnote}
and references therein)
is considered to be one of the powerful tool in the experiments 
at the LHC from the NP exploration standpoint.
The cause of this is high sensitivity of the dilepton continuum to any
modification of SM induced by NP.
The research  of the NP effects is impossible without the
exact knowledge of the SM predictions including higher-order
QCD and electroweak radiative corrections.

There have been numerous publications on this problem; let us now
determine the place of the presented paper (further we will
discuss only the electroweak corrections, the QCD contributions
are out of the presented paper). So, all of the electroweak
corrections to (\ref{1}) can be divided conventionally
into three categories:
\begin{itemize}
\item
(I) electroweak corrections induced by gauge Boson Self Energies (BSE),
\item
(II) the other QED corrections (i.e. radiative corrections
induced by at least one additional photon: virtual or real),
\item
(III) the other Weak Radiative Corrections (WRC) (i.e. radiative corrections
induced by additional heavy bosons: $Z$ or $W$),
\end{itemize}
The first and the second contributions have already been studied 
(see papers on
pure QED corrections \cite{MosShuSor}, and  the QED and electroweak
corrections in the
Z-peak region and above in \cite{DY2002},
and numerous papers cited there).
Naturally the contribution (II) also requires the consideration of 
the diagrams
with real bremsstrahlung (to cancel the infrared divergence) and,
therefore, causes the experimentally-motivated kinematical cuts
for observed photon; all of that has been taken into account 
in paper \cite{DY2002}.

To describe all contributions to electroweak corrections,
it is convenient to use the terminology
of the so-called electroweak  Sudakov logarithms \cite{sud-log},
i.e. the expressions which are growing with the scale of energy,
and thus giving one of the main effects in the region of large
invariant dilepton mass.
By now extensive studies have been done in this area.  
For instance, the weak Sudakov expansion for general 
four-fermion processes has been studied in detail
(see, for example, \cite{DENPOZ} 
and the recent paper \cite{ARX05} along with 
the extensive list of references therein).
Let us note that the (I) contribution
contains single Sudakov logarithms, and in the (II) contribution 
(in $Z\gamma$-boxes) there is also double Sudakov logarithms (DSL) 
but they are 
mutually cancelled out and reduced to single ones (see Section IV).
Finally, the (III) contribution contains the DSL,  
and as it will be shown below, they predominate in the region $M \gg m_Z$.
Besides, all of the contributions (I-III) contain zero power of Sudakov logs,
i.e. terms without weak boson masses.

Obviously, the collinear logarithms of (II) QED radiative
corrections can compete with the DSL in the investigated region.
This important issue has been studied
at the one-loop level in \cite{DY2002},
where both the QED and weak corrections have been calculated
for $M \leq$~2~TeV,
but it has yet remained unsolved in the region of $M>$~2~TeV.
Other important contributions in the investigated reaction
at high invariant masses
are the higher-order corrections (two-loop electroweak logarithms, at least),
which also have been studied in the works 
\cite{Denner:2006jr}, \cite{ARX05} 
(see also the numerous papers cited therein).
Weak boson emission contribution has been recently calculated 
in \cite{Baur2006}
and the contribution of higher-order corrections due to multiple photon 
emission has been computed in \cite{CarCal},
these contributions are beyond the presented calculations.

As has already been mentioned, for the future experiments at LHC
aimed at the searches of NP in the reaction (\ref{1})
it is important to know exactly the SM predictions,
including the radiative background, i.e. the processes,
which are experimentally indistinguishable from (\ref{1}).
The important task is the insertion of this background into the 
CMS Monte Carlo generators, which should be both accurate 
and fast. For them to be fast it is necessary to have a set of 
compact formulas for the WRC which have been obtained here.
Though we use the Asymptotic Approach (AA)
(the Sudakov logarithms as the parameters of expansion
in the investigated region of large $M$), 
retaining the first and zero powers
in the expansion we get good accuracy of calculation. 

Thus, we further introduce basic notations (Section II),
show how to calculate the cross section corresponding 
to the (III) and (I) contribution diagrams (see Fig.1,b-h):
the heavy vertices (Section III), the heavy boxes (Section IV),
and the boson self energies (Section V),
we compare our results with those available at the quark level,
and we give the numerical estimations of weak radiative corrections
to the Drell-Yan cross section
in the region $0.5\ \mbox{TeV} \leq M < 10 \leq \mbox{TeV}$ (Section~V).

\begin{figure*}
\vspace*{15mm}
\hspace*{-5mm}
\scalebox{0.25}{\includegraphics{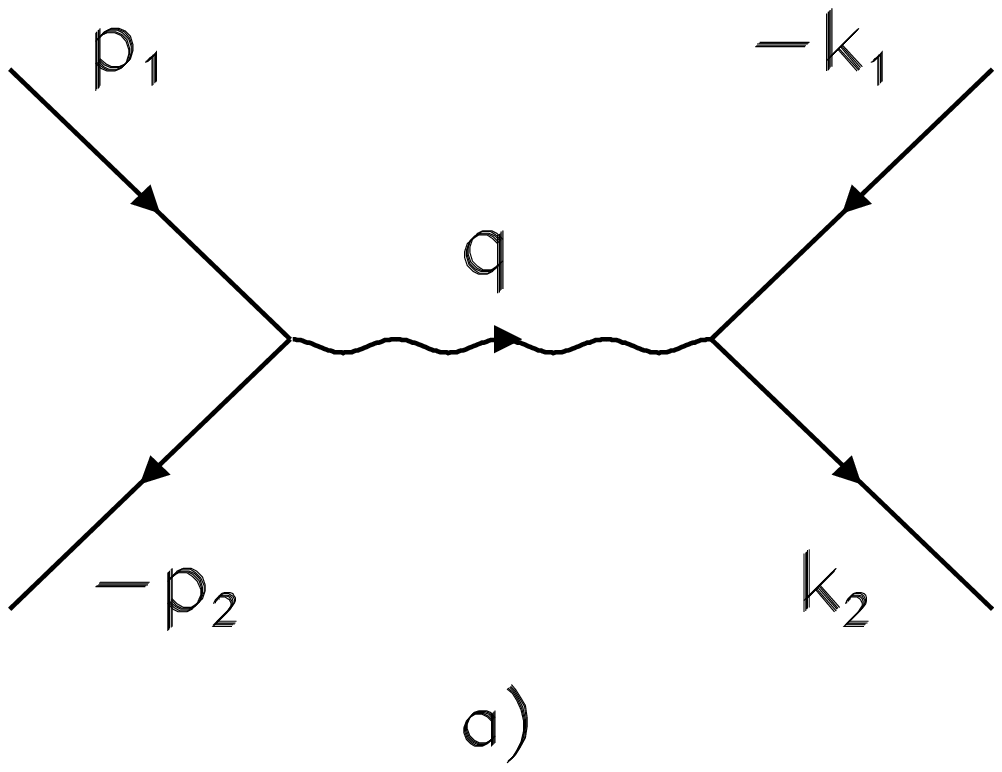}}
\scalebox{0.25}{\includegraphics{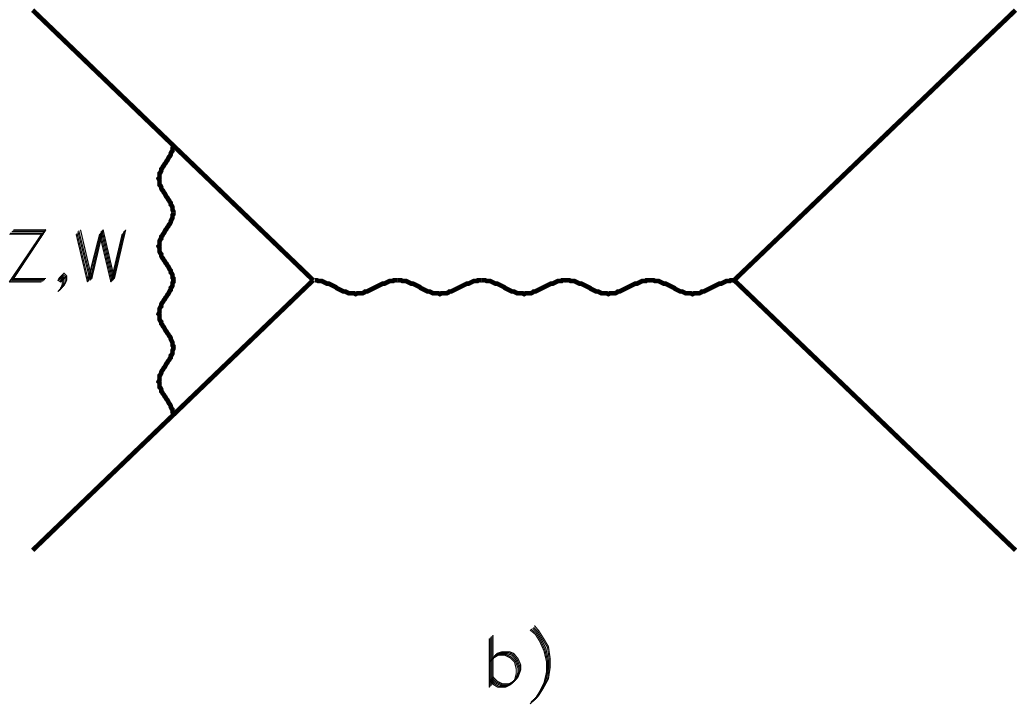}}
\scalebox{0.25}{\includegraphics{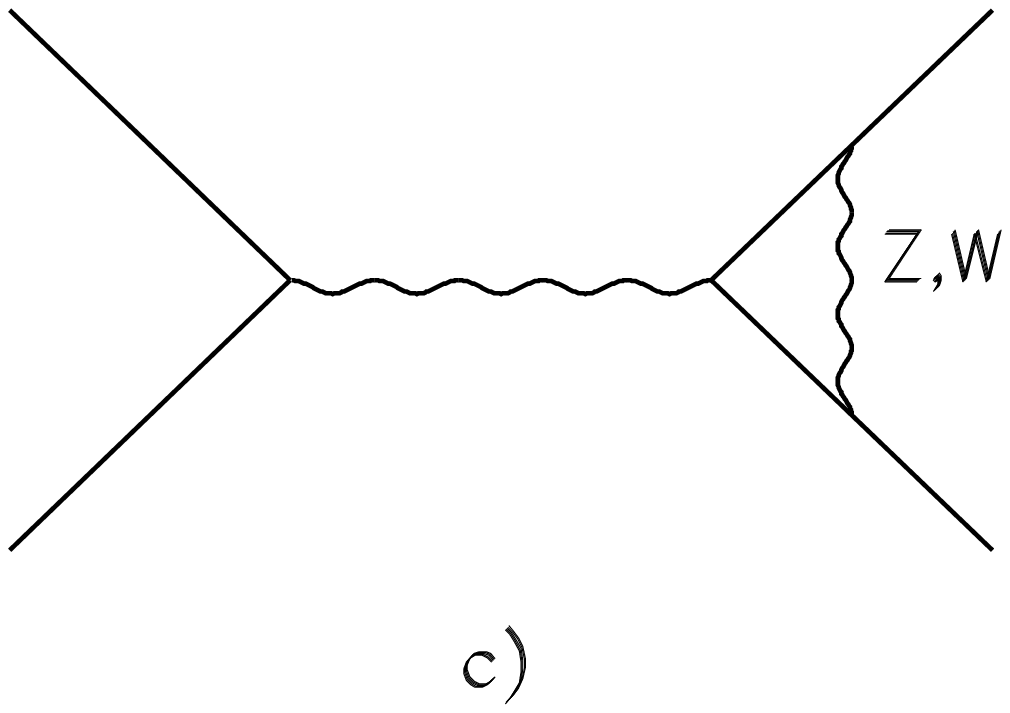}}
\scalebox{0.25}{\includegraphics{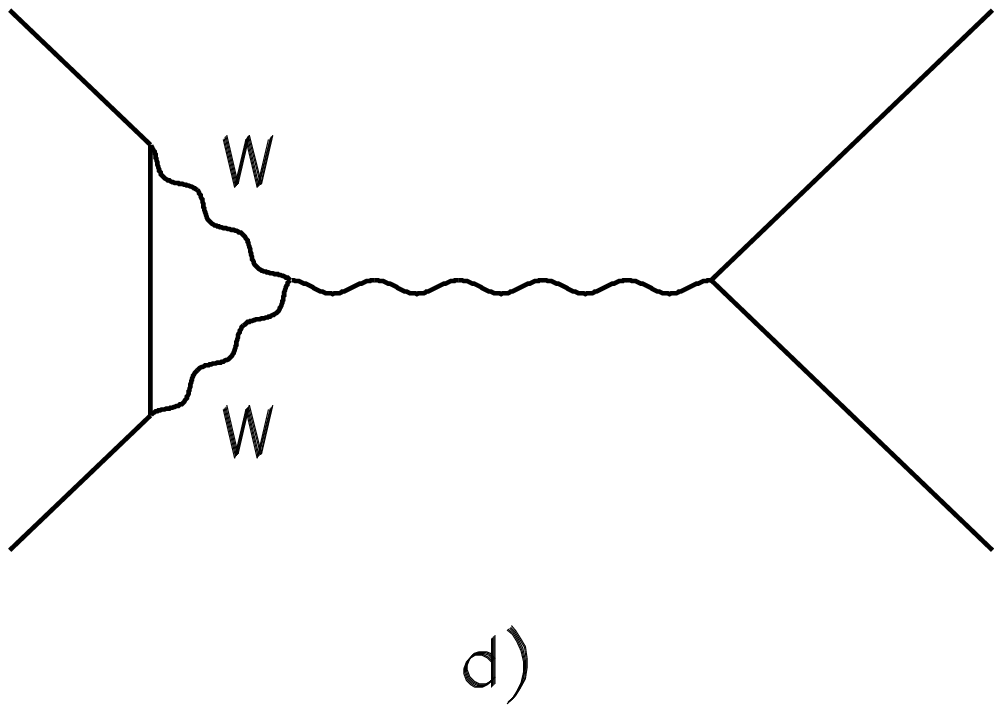}}
\scalebox{0.25}{\includegraphics{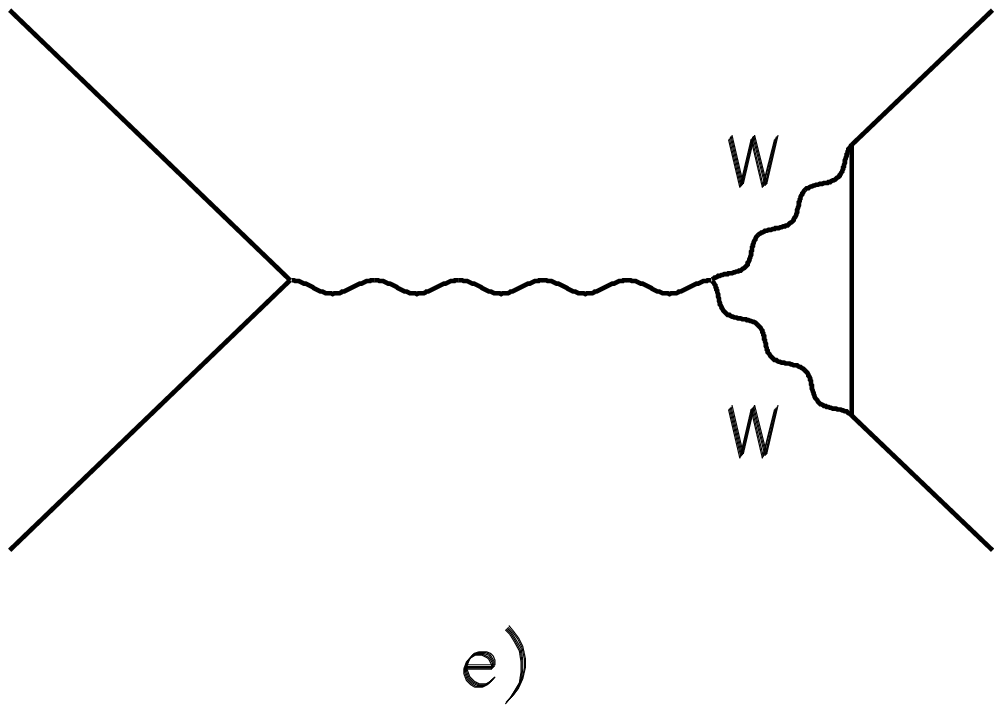}}
\hspace*{-15mm}
\scalebox{0.25}{\includegraphics{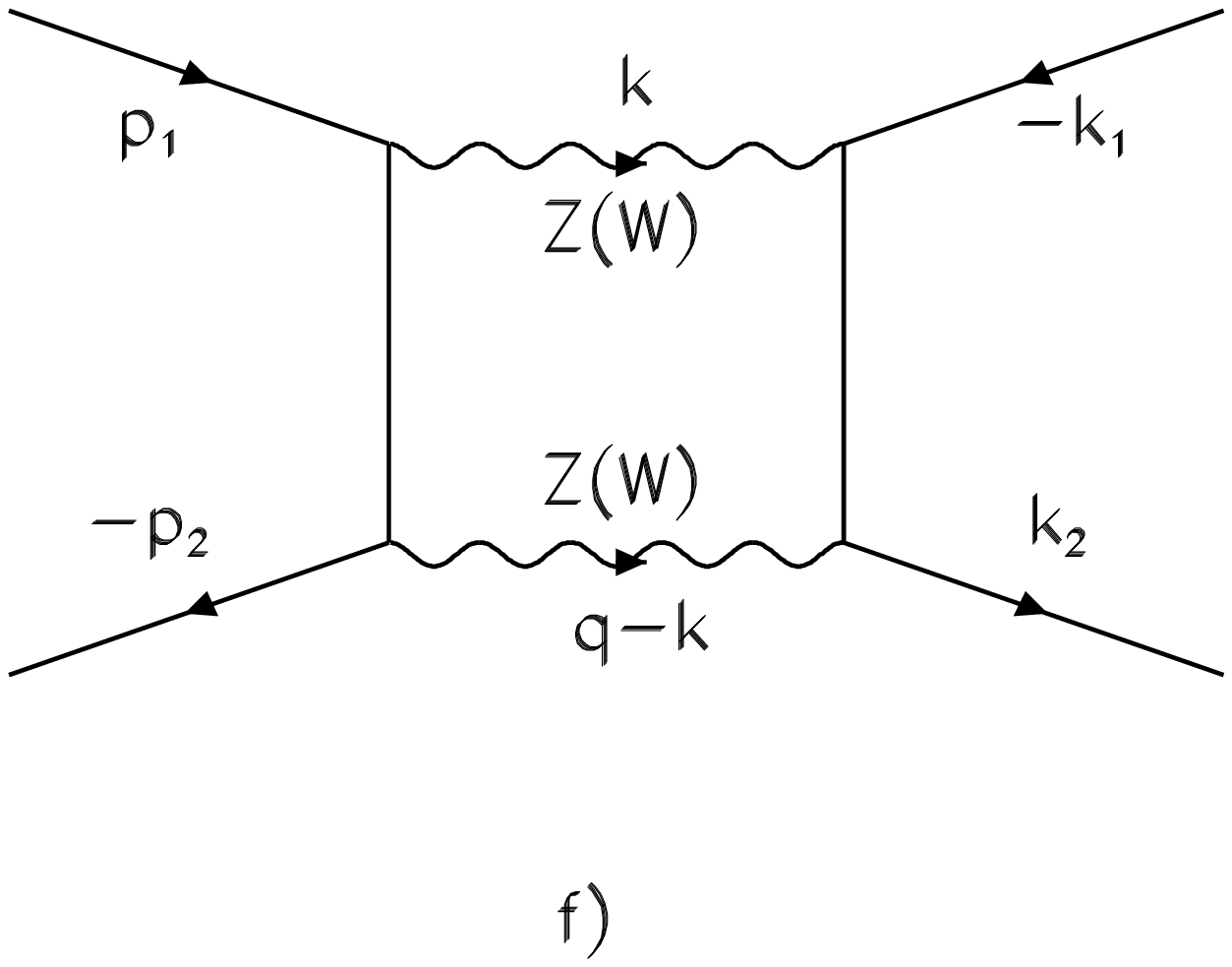}}
\scalebox{0.25}{\includegraphics{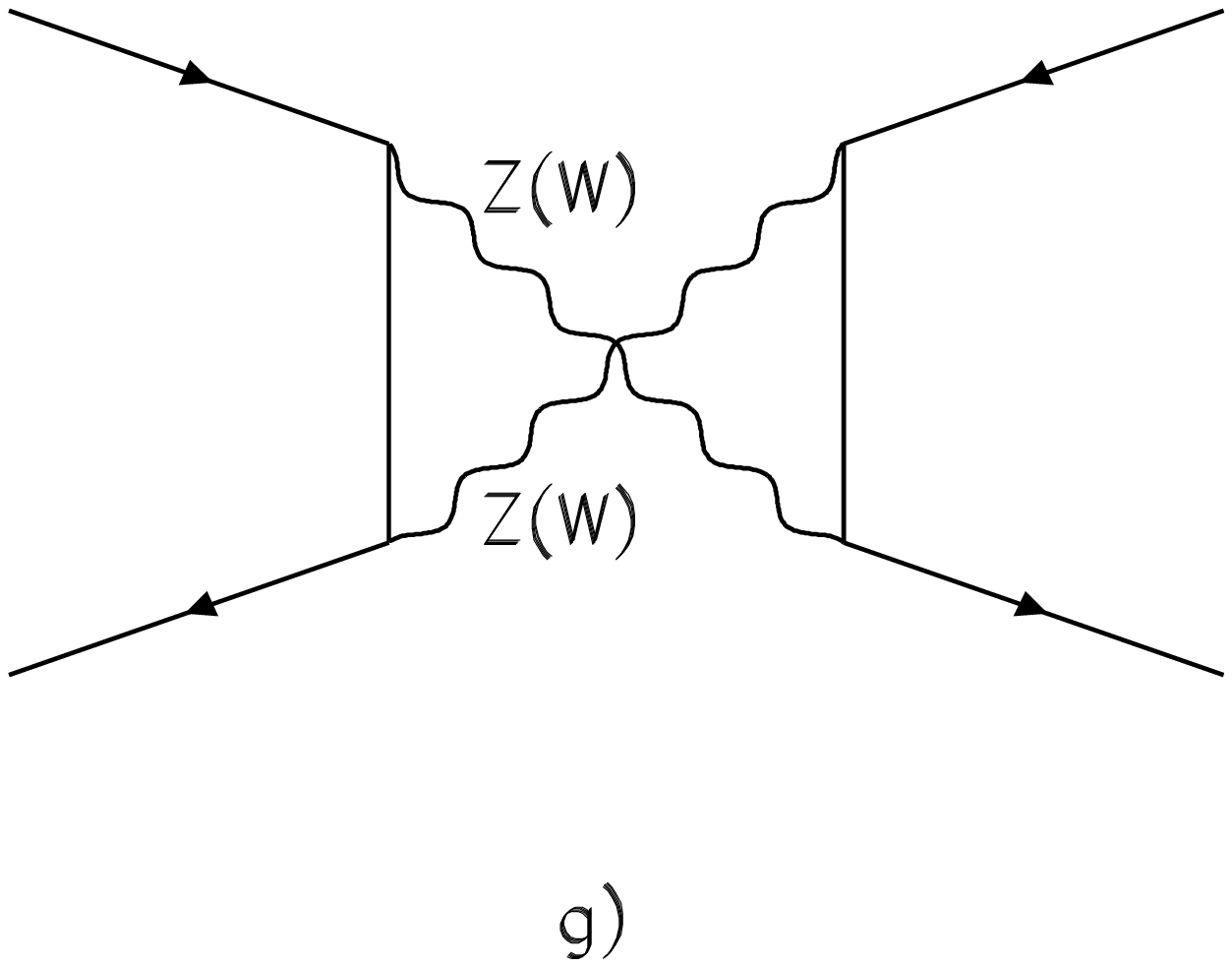}}
\hspace*{5mm}
\scalebox{0.25}{\includegraphics{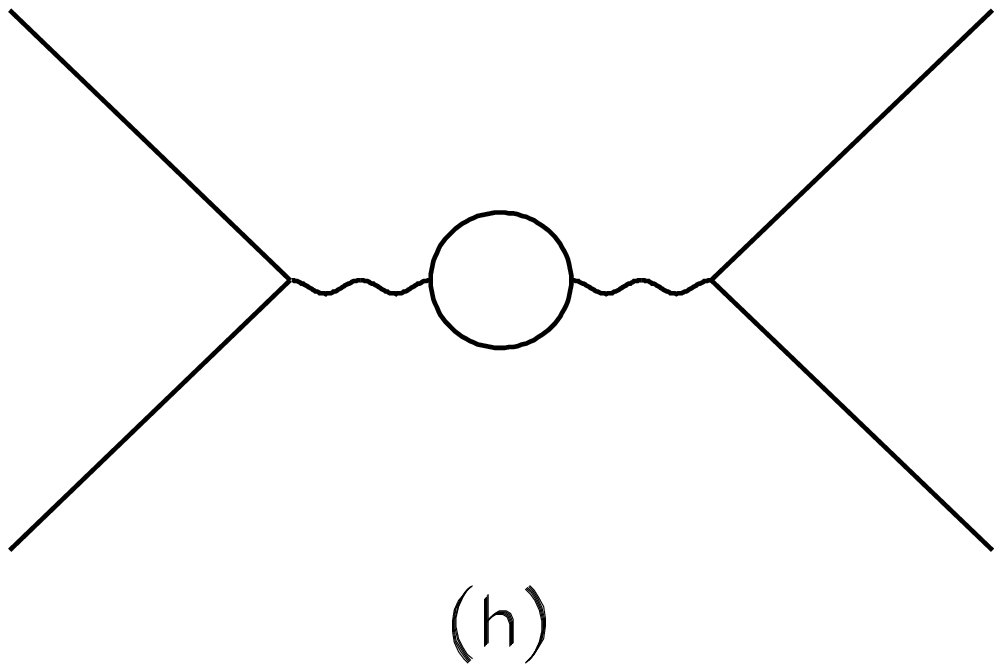}}
\vspace*{-25mm}
\caption{\label{fig:1}
Feynman graphs for the Born (a) and one-loop (b-h) diagrams with
additional virtual heavy bosons corresponding to the WRC.
Unsigned helix lines mean $\gamma$ or $Z$.
}
\end{figure*}

\section{Notations and the Born cross section of the Drell-Yan process}
The Born cross section for the inclusive hadronic reaction
$AB~\rightarrow~l^+l^-X $
is given in the quark parton model by formula
\begin{equation}
   \sigma_0(M,y,\zeta) =
   \frac{1}{3} \frac{2 \pi \alpha^2}{SM}
   {\mbox Re}
   \sum_{i,j=\gamma,Z} D^{is} {D^{js}}^*
   \sum_{\chi=+,-} (t^2+\chi u^2) {\lambda_l}^{i,j}_{\chi}
   \sum_{q=u,d,s,...} F_{\chi}^q(x_+,x_-) {\lambda_q}^{i,j}_{\chi}.
\label{fin-b}
\end{equation}
Our notations are the following (see Fig.1,a):
$p_1$ is the 4-momentum of the first unpolarized (anti)quark with the 
flavor $q$ and mass $m_q$;\
$p_2$ is the 4-momentum of the second (anti)quark with the 
flavor $q$ and mass $m_q$;\
$k_1 (k_2)$ is the 4-momentum of the final charged lepton $l^+ (l^-)$
with the  mass $m$;\ \
$q=k_1+k_2$ is the 4-momentum of the $i$-boson with the mass $m_i$;\ \
$P_{A(B)}$ is the 4-momentum of initial nucleons A(B).
The invariant mass of dilepton is $M=\sqrt{q^2}$.
We use the standard set of Mandelstam invariants for the partonic elastic
scattering $s,\ t,\ u$:
\begin{equation}
s=(p_1+p_2)^2,\ t=(p_1-k_1)^2,\ u=(k_1-p_2)^2
\end{equation}
and the invariant \ $S=(P_A+P_B)^2$  for hadron scattering.
The first summation runs over two possible intermediate bosons:
$\gamma, Z$.
The third summation runs over all contributing parton configurations.
The number $1/3$ is a color factor, $\alpha$ is the electromagnetic
fine structure constant.
The propagator for $j$-boson has the form
\begin{equation}
D^{js} =\frac{1}{s-m_j^2+im_j\Gamma_j},
\end{equation}
where $\Gamma_j$ is the $j$-boson width.
The combinations of parton density functions look like
\begin{equation}
  F_{\pm}^q(x_1,x_2) =    f_q^A(x_1)f_{\bar q}^B(x_2)
                     \pm  f_{\bar q}^A(x_1)f_q^B(x_2),
\end{equation}
where $f_q^H(x)$ is the probability of constituent $q$
with the fraction $x$ of the hadron's momentum in hadron $H$  finding.

The combinations of coupling constants for $f$-fermion with $i$- (or $j$-)
boson have the form
\begin{equation}
{\lambda_f}^{i,j}_+=v^i_fv^j_f+a^i_fa^j_f,\
{\lambda_f}^{i,j}_-=v^i_fa^j_f+a^i_fv^j_f,
\label{lamb}
\end{equation}
where
\begin{equation}
  v^{\gamma}_f=-Q_f,\
  a^{\gamma}_f=0,\
  v^Z_f=\frac{I_f^3-2s_W^2Q_f}
             {2s_Wc_W},\
  a^Z_f=\frac{I_f^3}{2s_Wc_W},
\end{equation}
$Q_f$ is the charge of fermion $f$,
$I_f^3$ is the third component of the weak isospin of fermion $f$, and
$s_W\ (c_W)$ is the sine(cosine) of the weak mixing angle:
$s_W=\sqrt{1-c_W^2}$, $c_W= m_W/m_Z$.

Reducing the phase space of reaction (\ref{1}) to the variables $M$ and
$y$ -- rapidity of dilepton
(we need also the variable $\tau$ which is determined by $\tau^2=q^2/S$)
we get  a three-, two- and one-fold cross sections
(without any experimental restrictions yet)
\begin{equation}
   \sigma(M,y,\zeta) \equiv \frac{d^3\sigma}{dM dy d\zeta},
\label{genn}
\end{equation}
\begin{equation}
   \sigma(M,y) \equiv \frac{d^2\sigma}{dM dy}
   = \int\limits_{-1}^{+1} d\zeta \ \sigma(M,y,\zeta),
\label{genn1}
\end{equation}
and
\begin{equation}
   \sigma(M) \equiv \frac{d\sigma}{dM}=
   \int\limits_{-\ln\frac{\sqrt{S}}{M}}^{+\ln\frac{\sqrt{S}}{M}} dy
\int\limits_{-1}^{+1}  d\zeta \ \sigma(M,y,\zeta),
\label{xsM}
\end{equation}
here the invariants $s$, $t$, and $u$ mean
\begin{equation}
s=M^2,\
t=-\frac{1}{2}s(1 - \zeta),\ u=-\frac{1}{2}s(1 + \zeta),
\end{equation}
$\zeta$ is cosine of the angle $\theta$
between $\vec p_1$ and $\vec k_1$ in the center
mass system of hadrons ($\zeta=\cos\theta$)
and the arguments of parton distribution functions
in (\ref{fin-b}) have the form
\begin{equation}
   x_{\pm} =  \tau e^{\pm y}.
\end{equation}
Integrating over the whole region of $\zeta$ 
we get much simpler expression
\begin{equation}
   \sigma_0(M,y) =
   \frac{8 \pi \alpha^2 M^3}{9S}
   {\mbox Re}
   \sum_{i,j=\gamma,Z} D^{is} {D^{js}}^*
   {\lambda_l}^{i,j}_{+}
   \sum_{q=u,d,s,...} F_{+}^q(x_+,x_-) {\lambda_q}^{i,j}_{+}.
\end{equation}

\section{Heavy Vertices}
To construct the contribution of Heavy Vertices (HV) 
to the radiative corrections
(see diagrams (b-e) in Fig.1), we used
the t'Hooft-Feynman gauge and an on-mass renormalization scheme which uses
$\alpha, m_W, m_Z,$ Higgs boson mass $m_H$, and the fermion masses
as independent parameters.
Appropriate (for ultrarelativistic limit we are interested in)
results can be taken from Ref.\cite{BSH86},
where renormalized gauge boson fermion vertices for on
shell fermions have been obtained.
These results are presented as the form factor set to the Born vertices,
so we can easily use them to construct the cross section:
hence we replace the coupling constants in the Born vertex
for the corresponding form factors:
\begin{equation}
v_f^j \rightarrow  F_V^{jf},
a_f^j \rightarrow  F_A^{jf}.
\end{equation}
Then the cross section of heavy vertices contribution
will look like
\begin{eqnarray}
   \sigma_V(M,y,\zeta) =
   \frac{4 \pi \alpha^2}{3SM}
   &&{\mbox Re}
   \sum_{i,j=\gamma,Z} D^{is} {D^{js}}^*
   \sum_{\chi=+,-} (t^2+\chi u^2) 
\sum_{q=u,d,s,...} F_{\chi}^q(x_+,x_-)
({\lambda_q^F}^{i,j}_{\chi} {\lambda_l}^{i,j}_{\chi} +
 {\lambda_q}^{i,j}_{\chi} {\lambda_l^F}^{i,j}_{\chi}),
\label{fin-V}
\end{eqnarray}
where the combinations of couplings constants and form factors are
\begin{eqnarray}
{\lambda_f^F}^{i,j}_{+} =  F_V^{if} v_f^j +  F_A^{if} a_f^j,\
{\lambda_f^F}^{i,j}_{-} =  F_V^{if} a_f^j +  F_A^{if} v_f^j,\ \
(f=q,l).
\end{eqnarray}

Electroweak form factors $ F_{V,A}^{if}$ in ultrarelativistic
limit have the form (there is the perfect coincidence with
the formulas of Appendix C of Ref.\cite{Hollik}):
\begin{eqnarray}
F_V^{\gamma l} & = &
  \frac{\alpha v_l^{\gamma}}{4\pi}
  [ ({(v_l^Z)}^2 + {(a_l^Z)}^2 ) \Lambda_2(m_Z)
   + \frac{3}{4s_W^2} \Lambda_3(m_W) ],
\\[0.3cm] \displaystyle
F_A^{\gamma l} & = &
  \frac{\alpha v_l^{\gamma}}{4\pi}
  [ 2v_l^Za_l^Z \Lambda_2(m_Z)
   + \frac{3}{4s_W^2} \Lambda_3(m_W) ],
\nonumber \\[0.3cm] \displaystyle
F_V^{\gamma d} & = &
  \frac{\alpha v_d^{\gamma} }{4\pi}
  [ ({(v_d^Z)}^2 + {(a_d^Z)}^2 ) \Lambda_2(m_Z)
   - \frac{1}{2s_W^2} \Lambda_2(m_W)
   + \frac{9}{4s_W^2} \Lambda_3(m_W) ],
\nonumber \\[0.3cm] \displaystyle
F_A^{\gamma d} & = &
  \frac{\alpha v_d^{\gamma} }{4\pi}
  [ 2v_d^Za_d^Z \Lambda_2(m_Z)
   - \frac{1}{2s_W^2} \Lambda_2(m_W)
   + \frac{9}{4s_W^2} \Lambda_3(m_W) ],
\nonumber \\[0.3cm] \displaystyle
F_V^{\gamma u} & = &
  \frac{\alpha v_u^{\gamma} }{4\pi}
  [ ({(v_u^Z)}^2 + {(a_u^Z)}^2 ) \Lambda_2(m_Z)
   - \frac{1}{8s_W^2} \Lambda_2(m_W)
   + \frac{9}{8s_W^2} \Lambda_3(m_W) ],
\nonumber \\[0.3cm] \displaystyle
F_A^{\gamma u} & = &
  \frac{\alpha v_u^{\gamma} }{4\pi}
  [ 2v_u^Za_u^Z \Lambda_2(m_Z)
   - \frac{1}{8s_W^2} \Lambda_2(m_W)
   + \frac{9}{8s_W^2} \Lambda_3(m_W) ].
\nonumber
\end{eqnarray}
\begin{eqnarray}
F_V^{Z l} & = &
  \frac{\alpha }{4\pi}
  [  v_l^Z ({(v_l^Z)}^2 + 3{(a_l^Z)}^2 ) \Lambda_2(m_Z)
   + \frac{1}{8s_W^3c_W} \Lambda_2(m_W)
   - \frac{3c_W}{4s_W^3} \Lambda_3(m_W) ],
\\[0.3cm] \displaystyle
F_A^{Z l} & = &
  \frac{\alpha }{4\pi}
  [ a_l^Z (3{(v_l^Z)}^2 + {(a_l^Z)}^2 ) \Lambda_2(m_Z)
   + \frac{1}{8s_W^3c_W} \Lambda_2(m_W)
   - \frac{3c_W}{4s_W^3} \Lambda_3(m_W) ],
\nonumber \\[0.3cm] \displaystyle
F_V^{Z d} & = &
  \frac{\alpha }{4\pi}
  [ v_d^Z ({(v_d^Z)}^2 + 3{(a_d^Z)}^2 ) \Lambda_2(m_Z)
   + \frac{1-2Q_us_W^2}{8s_W^3c_W} \Lambda_2(m_W)
   - \frac{3c_W}{4s_W^3} \Lambda_3(m_W) ],
\nonumber \\[0.3cm] \displaystyle
F_A^{Z d} & = &
  \frac{\alpha }{4\pi}
  [ a_d^Z (3{(v_d^Z)}^2 + {(a_d^Z)}^2 ) \Lambda_2(m_Z)
   + \frac{1-2Q_us_W^2}{8s_W^3c_W} \Lambda_2(m_W)
   - \frac{3c_W}{4s_W^3} \Lambda_3(m_W) ],
\nonumber \\[0.3cm] \displaystyle
F_V^{Z u} & = &
  \frac{\alpha }{4\pi}
  [ v_u^Z ({(v_u^Z)}^2 + 3{(a_u^Z)}^2 ) \Lambda_2(m_Z)
   - \frac{1+2Q_ds_W^2}{8s_W^3c_W} \Lambda_2(m_W)
   + \frac{3c_W}{4s_W^3} \Lambda_3(m_W) ],
\nonumber \\[0.3cm] \displaystyle
F_A^{Z u} & = &
  \frac{\alpha }{4\pi}
  [  a_u^Z (3{(v_u^Z)}^2 + {(a_u^Z)}^2 ) \Lambda_2(m_Z)
   - \frac{1+2Q_ds_W^2}{8s_W^3c_W} \Lambda_2(m_W)
   + \frac{3c_W}{4s_W^3} \Lambda_3(m_W) ].
\nonumber
\end{eqnarray}

Here it is necessary to note that in on-mass renormalization scheme
the self-energies of $u$-quarks' diagrams also give a non-zero
contribution to the  cross section. However, the results for
$u$-self-energy are factorized in the same manner as the vertices.
It gives us a possibility to sum both contributions in one formula,
where the $u$-self-energy contribution is completely cancelled with
the corresponding terms of heavy vertices.
So, it could be said that the formula (\ref{fin-V}) gives the cross section
induced by both the heavy vertices and self-energy of $u$-quarks.

Exact expressions for $\Lambda_{2,3}$ can be found in \cite{BSH86}, 
and there
we give the real part of expressions for functions $\Lambda_{2,3}(m_i)$
that provide a good accuracy in large dilepton mass region
(see Introduction):
\begin{equation}
\Lambda_2(m_i)=\frac{\pi^2}{3}-\frac{7}{2}-3l_{i,s}-l_{i,s}^2,\ \ \
\Lambda_3(m_i)=\frac{5}{6}-\frac{1}{3}l_{i,s}.
\label{L23}
\end{equation}
The Sudakov logarithm that we denote as $l_{i,x}$ has the form
\begin{equation}
l_{i,x} = \log\frac{m_i^2}{|x|}\ \ \ \ (i=Z,W;\ \  x=s,t,u).
\label{sud-log}
\end{equation}

\section{Heavy Boxes}
The calculation of a two heavy boson (box) contribution is more complicated
procedure since it requires the integration of 4-point functions
with the complex masses in unlimited from above kinematical region of
invariants. 
The difficulties of such a procedure have been pointed out
as far back as a pioneer paper, Ref. \cite{HooftVeltman}.
Fortunately there is a way to avoid many troubles with the
integration all of the terms in the box contribution using a rather simple
algebra. Let us explain this way.

First of all we construct the ZZ-box cross section for
$q\bar q \rightarrow l^+l^-$ (see diagrams in Fig.1,f-g)
using the standard Feynman rules:
\begin{equation}
   d\sigma_{\rm ZZ} = -\frac{4 \alpha^3}{\pi s}
  d\Gamma_q  {\mbox Re}   \frac{i}{(2\pi)^2} \int d^4k
   \sum_{k=\gamma,Z} {D^{ks}}^* (D^{\rm ZZ}+C^{\rm ZZ}),
\label{zz}
\end{equation}
here the 2-particle phase space element is read 
$$d\Gamma_q=
  \frac{d^3k_1}{2k_1^0} \frac{d^3k_2}{2k_2^0}
  \delta(q-p_1-p_2),$$
and $D^{\rm ZZ}(C^{\rm ZZ})$ 
is the direct (crossed) diagram contribution.

Neglecting the fermion masses
we present the direct contribution in the form:
\begin{equation}
D^{\rm ZZ}=\Pi^D_{\rm ZZ}
\mbox{Tr}
[\gamma^{\alpha}\hat p_2\gamma_\mu(\hat p_1-\hat k)
\gamma_\nu \rho_q^{\rm ZZ,k}(p_1)]
\mbox{Tr}
[\gamma_{\alpha}\hat k_2\gamma^\mu(\hat k-\hat k_1)
\gamma^\nu \rho_l^{\rm ZZ,k}(k_1)],
\label{D}
\end{equation}
where the integrand of 4-point scalar function looks like
\begin{equation}
\Pi^D_{\rm ZZ}=\frac{1}{((q-k)^2-m_Z^2)(k^2-m_Z^2)(k^2-2k_1k)(k^2-2p_1k)}.
\label{pi4}
\end{equation}
Combinations of the density matrices $\rho(p)$ and the coupling constants
can be reduced to the production of $\lambda$-factors 
(see formulas (\ref{lamb})) and
$\hat p \equiv \gamma_\mu p^\mu$ (here $p = p_1,p_2,k_1,k_2$ or $k$)
\begin{equation}
\rho_f^{\rm ZZ,k}(p)=(v_f^{\rm ZZ}-a_f^{\rm ZZ}\gamma_5)\rho(p)(v_f^{k}+a_f^{k}\gamma_5)
=\frac{1}{2}({\lambda_f}^{\rm ZZ,k}_+ - {\lambda_f}^{\rm ZZ,k}_-\gamma_5)\hat p,
\label{ro}
\end{equation}
\begin{equation}
v_f^{\rm ZZ}={(v_f^Z)}^2+{(a_f^Z)}^2,\
a_f^{\rm ZZ}=2v_f^Za_f^Z.
\label{vzz}
\end{equation}

To extract the part of cross section which predominates in the region
$|x| \gg m_Z^2$ (see $x$ in formula (\ref{sud-log})) we should make
the equivalent transformation of the cross section based on the close 
connection
of infrared divergency cross section terms and the Sudakov leading-log terms:
\begin{equation}
D^{\rm ZZ}=(D^{\rm ZZ}|_{k\rightarrow 0}+D^{\rm ZZ}|_{k\rightarrow q})+
       (D^{\rm ZZ}-D^{\rm ZZ}|_{k\rightarrow 0}-D^{\rm ZZ}|_{k\rightarrow q})
      =D_1^{\rm ZZ} + D_2^{\rm ZZ}.
\label{equi}
\end{equation}
The integral over $k$ of the first term in (\ref{equi}) is
\begin{eqnarray}
\frac{i}{(2\pi)^2} 
\int d^4k D_1^{\rm ZZ}
= \frac{4t}{q^2-m_Z^2}  B^{\rm ZZ,k}
\frac{i}{(2\pi)^2} 
\int  
\frac{d^4k}{k^2-m_Z^2}
\Bigl[ \frac{1}{(k^2-2k_1k)(k^2-2p_1k)} +
 \frac{1}{(k^2-2k_2k)(k^2-2p_2k)} \Bigr],
\label{D1}
\end{eqnarray}
here we used the trivial correlation typical for the Born cross section
\begin{equation}
B^{\rm ZZ,k}
=
\frac{1}{2}
\mbox{Tr}
[\gamma^{\alpha}\hat p_2\gamma_\mu
\rho_q^{\rm ZZ,k}(p_1)]
\mbox{Tr}
[\gamma_{\alpha}\hat k_2\gamma^\mu
\rho_l^{\rm ZZ,k}(k_1)]
\approx
b_+^{\rm ZZ,k}t^2+b_-^{\rm ZZ,k}u^2,
\end{equation}
where
\begin{equation}
b_{\pm}^{n,k}=
{\lambda_q}^{n,k}_+{\lambda_l}^{n,k}_+
\pm {\lambda_q}^{n,k}_-{\lambda_l}^{n,k}_-.
\label{bplusminus}
\end{equation}

Then we integrate one of terms in (\ref{D1})  over $k$ using the standard
method of paper \cite{HooftVeltman} (another term is integrated similarly):
\begin{eqnarray}
&&\frac{i}{(2\pi)^2}  \int
\frac{d^4k }{(k^2-m_Z^2)(k^2-2k_2k)(k^2-2p_2k)}
= \frac{1}{4} \int\limits_0^1 dx \int\limits_0^x
\frac{dy}{m_Z^2(x-y)+(p_2(1-x)+k_2y)^2} =
\nonumber \\&&
= \frac{1}{4} \int\limits_0^1 dx
\frac{1}{t(x-1)-m_Z^2}\log \frac{t(x-1)}{m_Z^2}
= - \frac{1}{4t}
(\frac{\pi^2}{3}+\frac{1}{2}\log^2\frac{-t-m_Z^2}{m_Z^2}
+\mbox{Li}_2\frac{m_Z^2}{m_Z^2+t} ),
\end{eqnarray}
here $\mbox{Li}_2$ denotes the Spence dilogarithm.
Retaining the terms which are proportional to
the second ($ \sim l^2_{i,x}$),
first ($ \sim l^1_{i,x}$) and zero
($ \sim l^0_{i,x}=1$) power of the Sudakov logarithms
and thereby neglecting the terms which are inessential in the region
of large dilepton mass $|x| \gg m_Z^2$
we get the asymptotic expression
\begin{eqnarray}
\frac{i}{(2\pi)^2} \int d^4k D_1^{\rm ZZ}
\approx -\frac{1}{s} B^{\rm ZZ,k} (\frac{2\pi^2}{3} + l^2_{Z,t}).
\label{d1}
\end{eqnarray}

Surely in these (and subsequent) expressions we can retain only
leading $ \sim l^2_{i,x}$ term as it has been done, for example, 
in \cite{ital}.
Naturally in such an approximation our results coincide
with the results of that paper.
However we are able to retain the first and zero powers
of expansion leading-log parameter and we will use this opportunity
to improve the accuracy of radiative correction estimation
(one remark more  -- we retain the $ l^1_{i,x}$ and
$ l^0_{i,x}$ also in the heavy vertices part).

The asymptotic expression (at  $|x| \gg m_Z^2$) for the second term
of the formula (\ref{equi}) can be presented as
\begin{equation}
D_2^{\rm ZZ} \approx
D^{\rm ZZ}|_{\Pi^D_{\rm ZZ} \rightarrow \Pi^D_{\gamma \gamma}}
- 4 \Pi^D_{\gamma \gamma} \frac{(q-k)^2+k^2}{q^2}tB^{\rm ZZ,k}.
\label{d2}
\end{equation}
Reducing the vectorial and tensor integrals to the scalar ones
we get
\begin{eqnarray}
\frac{i}{(2\pi)^2} \int d^4k D_2^{\rm ZZ} \approx &&
- 2 b_-^{\rm ZZ,k}[(G_0^m+G_0^M)(t^2+u^2)-2(R-N_0)u-X_0t\frac{t^2+u^2}{s}] -
4 b_+^{\rm ZZ,k}t^2[G_0^m+G_0^M-X_0\frac{t}{s}],
\end{eqnarray}
where scalar integrals are
\begin{eqnarray}
G_0^{m} &=& \frac{i}{(2\pi)^2} \int
\frac{d^4k}{(k^2-2k_1k)k^2(k-q)^2}
=-\frac{1}{8q^2}(\log^2\frac{q^2}{m^2}+\frac{\pi^2}{3}),
\nonumber \\[0.3cm] \displaystyle
G_0^{M} &=& \frac{i}{(2\pi)^2} \int
\frac{d^4k}{(k^2-2p_1k)k^2(k-q)^2}
=-\frac{1}{8q^2}(\log^2\frac{q^2}{m_q^2}+\frac{\pi^2}{3}),
\nonumber \\[0.3cm] \displaystyle
R &=& \frac{i}{(2\pi)^2} \int
\frac{d^4k}{k^2(k-q)^2}
=\frac{1}{4}(\log\frac{q^2}{L}-1),
\nonumber \\[0.3cm] \displaystyle
N_0 &=& \frac{i}{(2\pi)^2} \int
\frac{d^4k}{(k^2-2k_1k)(k^2-2p_1k)}
=\frac{1}{4}(\log\frac{m^2}{L}-1-\log\frac{m}{m_q}+\log\frac{|2k_1p_1|}{mm_q}),
\nonumber \\[0.3cm] \displaystyle
X_0 &=& \frac{i}{(2\pi)^2} \int
d^4k \frac{2k(q-k)}{(k^2-2k_1k)(k^2-2p_1k)k^2(k-q)^2}=
\nonumber \\[0.3cm] \displaystyle
&=&-\frac{1}{8p_1k_1}(
\frac{1}{2} \log\frac{|2p_1k_1|m^2}{q^4} \log\frac{|2p_1k_1|}{m^2}
+\frac{1}{2} \log\frac{|2p_1k_1|m_q^2}{q^4} \log\frac{|2p_1k_1|}{m_q^2}
-\frac{\pi^2}{3}).
\label{scal}
\end{eqnarray}
These integrals have been calculated on the analogy with the corresponding
ones in paper of J.Kahane \cite{Kahane}
(with the important distinctions due to
different reaction channel: $t$-channel in \cite{Kahane}, and
$s$-channel in this paper). The expressions (\ref{scal})
contain the parameters of two sorts: masses of fermions and
$L$ -- parameter regulating the ultraviolet divergence, both of
them are completely cancelled out in the final expression
\begin{eqnarray}
\frac{i}{(2\pi)^2} \int d^4k D_2^{\rm ZZ} \approx
b_-^{\rm ZZ,k} l_s u
+ (b_-^{\rm ZZ,k} (t^2+u^2) + 2 b_+^{\rm ZZ,k}t^2 ) \frac{1}{2s} l_s^2,\ \
l_s=\log\frac{s}{|t|}.
\end{eqnarray}

The contribution of crossed part (Fig.1,g) to the cross section
can be calculated on the analogy with the direct one.
Besides, there is an interesting symmetry between the direct and crossed
parts (see, for example, \cite{DY2002}), namely
\begin{eqnarray}
C^{\rm ZZ}=-D^{\rm ZZ}|_{t \leftrightarrow u,\ b_+^{\rm ZZ,k} \leftrightarrow b_-^{\rm ZZ,k}}.
\label{sym}
\end{eqnarray}

And now let us present the total final expression for the contribution 
of ZZ-box
into cross section of the Drell-Yan process at large invariant dilepton mass:
\begin{eqnarray}
   \sigma_{\rm ZZ}(M,y,\zeta) = 
   \frac{2 \alpha^3}{3SM}
   {\mbox Re}
   \sum_{k=\gamma,Z} {D^{ks}}^*  
&& 
\!\! \sum_{q=u,d,s,...}
\Bigl[ f_q^A(x_+) f_{\bar q}^B(x_-)
 \bigl(\delta^{\rm ZZ,k}(t,u,b_+,b_-) - \delta^{\rm ZZ,k}(u,t,b_-,b_+) \bigr) +
\nonumber \\&&
\ \ \ \ \ \ \ \ \  \ \ \
+      f_{\bar q}^A(x_+) f_{q}^B(x_-)
 \bigl(\delta^{\rm ZZ,k}(u,t,b_+,b_-) - \delta^{\rm ZZ,k}(t,u,b_-,b_+) \bigr)\Bigr],
\label{fin-zz}
\end{eqnarray}
where
\begin{eqnarray}
\delta^{\rm ZZ,k}(t,u,b_+,b_-)
=\frac{ B^{\rm ZZ,k}}{s}(\frac{2\pi^2}{3}+l^2_{Z,t})
- b_-^{\rm ZZ,k}ul_s
- (b_-^{\rm ZZ,k} (t^2+u^2) + 2 b_+^{\rm ZZ,k}t^2 ) \frac{l_s^2}{2s}.
\label{deltazz}
\end{eqnarray}
Integration over $\zeta$ of the formula (\ref{deltazz}) gives:
\begin{eqnarray}
\int\limits_{-1}^{+1}\delta^{\rm ZZ,k}(t,u,b_+,b_-)\ d\zeta &&\ =
\int\limits_{-1}^{+1}\delta^{\rm ZZ,k}(u,t,b_+,b_-)\ d\zeta\ =
\nonumber \\&&
\!\!\!\!\!\!\!\!\!\!\!\!\!\!\! =\frac{s}{9}
( 12{\lambda_q}^{\rm ZZ,k}_+{\lambda_l}^{\rm ZZ,k}_+(\frac{2\pi^2}{3}+l^2_{Z,s})
+(22b_-^{\rm ZZ,k}+4b_+^{\rm ZZ,k})l_{Z,s} + 27b_-^{\rm ZZ,k})
\label{ideltazz}
\end{eqnarray}
(it is convenient to use it for the construction of twofold cross sections
$\sigma_{\rm ZZ,WW}(M,y)$).

To obtain the  WW-box contribution into the Drell-Yan cross section
using the expressions (\ref{fin-zz}) and  (\ref{deltazz}) one should:
1) do the trivial substitution $Z \rightarrow W$
in all indices of coupling constants and boson masses,
2) take into consideration that some quark diagrams with two W-bosons
are forbidden by the charge conservation law.
Then WW-box contribution will look like
\begin{eqnarray}
   \sigma_{\rm WW}(M,y,\zeta) = 
   \frac{2 \alpha^3}{3SM}
   {\mbox Re}
&& 
   \sum_{k=\gamma,Z} {D^{ks}}^*  
\Bigl( \!\! \sum_{q=u,c}
\Bigl[ f_q^A(x_+) f_{\bar q}^B(x_-) \delta^{\rm WW,k}(t,u,b_+,b_-) 
 +     f_{\bar q}^A(x_+) f_{q}^B(x_-) \delta^{\rm WW,k}(u,t,b_+,b_-) \Bigr] -
\nonumber \\&&
\!\! - \!\! \sum_{q=d,s,b}
 \Bigl[ f_q^A(x_+) f_{\bar q}^B(x_-) \delta^{\rm WW,k}(u,t,b_-,b_+) 
 +     f_{\bar q}^A(x_+) f_{q}^B(x_-) \delta^{\rm WW,k}(t,u,b_-,b_+) \Bigr]
	\Bigl).
\label{fin-ww}
\end{eqnarray}
This second feature of WW-boxes explains the domination
of WW-contribution into the Drell-Yan cross section in comparison to
ZZ-contribution (see the section on numerical analysis).
The point is that the ZZ-box cross section is proportional to the difference
\begin{eqnarray}
\delta^{\rm ZZ,k}(t,u,b_+,b_-) - \delta^{\rm ZZ,k}(u,t,b_-,b_+)
\sim l^2_{Z,t} - l^2_{Z,u}
= \log\frac{u}{t}(l^1_{Z,t} + l^1_{Z,u}),
\label{diff}
\end{eqnarray}
(see also \cite{ital}) i.e. it $\sim l^1_{Z,x}$,
whereas the leading parts of WW--cross section do not
contain the difference (\ref{diff})
and are proportional to $l^2_{W,x}$.
Let us note here that the factorization property (\ref{diff})
is absent in heavy vertex part and is present in
infrared finite part of $\gamma Z$--box contribution, which
can be easily obtained on analogy with ZZ-box using
\begin{eqnarray}
\delta^{\gamma Z,k}(t,u,b_+,b_-)
=\frac{ B^{\gamma Z,k}}{s}(\frac{2\pi^2}{3}+l^2_{Z,t})
- 2b_-^{\gamma Z,k}ul_s
- (b_-^{\gamma Z,k} (t^2+u^2) + 2 b_+^{\gamma Z,k}t^2 ) \frac{l_s^2}{s}.
\label{deltagz}
\end{eqnarray}
We attribute the $\gamma Z$--box to the "other QED corrections"
(see (II) contribution in Introduction) and do not consider it here any more.

\section{Discussion of numerical results}
\subsection{Numerical estimation at the parton level and comparison with
the available results}
Now we discuss the observable quantities at the parton level.
The reason of it is twofold: to study the characteristics of the WRC
in the absence of parton distribution functions and
(what is much more important)
to compare them with the available results.
Except our paper \cite{YAFDY}, where 
the WRC to Drell-Yan production have been calculated 
by different  methods: exact integration, numerical
integration over Feynman parameters \cite{Fey} and the AA considered here,
there is the pioneer paper \cite{DY2002} and the calculation
of SANC group \cite{SANC}.

At the parton level there is a possibility to compare 
the results with \cite{DY2002},
since in this paper the set of relative non-QED corrections 
has been presented at high parton center
of mass energies for the total cross section (FIG. 3 of \cite{DY2002}) 
and for the parton forward-backward asymmetry (FIG. 4 of \cite{DY2002}).
Further we present
our variant of corresponding WRC in the \cite{DY2002} region (up to 1~TeV)
and in the region of higher energies (up to 10~TeV).
For the correct and tuned comparison we used in this subsection
the same set of the SM parameters
which was used in  \cite{DY2002} and all the rest prescriptions of this paper.
We also add the contribution of boson self energies 
to the numerical estimation (see Fig.1,h)
\begin{eqnarray}
d\sigma^{q \bar q}_{\rm BSE}=-\frac{8\alpha^2}{s} d\Gamma_q \bigl[ &&
\sum\limits_{i,j=\gamma,Z}
\Pi^i D^{is} {D^{js}}^* \sum\limits_{\chi=+,-}
{\lambda_q}^{i,j}_{\chi} {\lambda_l}^{i,j}_{\chi} (t^2+{\chi}u^2) +
\nonumber \\&&
+  \Pi^{\gamma Z} D^{Zs} \sum\limits_{i=\gamma,Z}
{D^{js}}^* \sum\limits_{\chi=+,-}
( {\lambda_q}^{\gamma,j}_{\chi} {\lambda_l}^{Z,j}_{\chi} +
{\lambda_q}^{Z,j}_{\chi} {\lambda_l}^{\gamma,j}_{\chi}) (t^2+{\chi}u^2) \bigr].
\label{bse}
\end{eqnarray}
Here
$\Pi^{\gamma,Z,\gamma Z}$ are connected with 
the renormalized photon--, Z-- and $\gamma$Z--self energies \cite{BSH86} as
$$ 
\Pi^{\gamma}=\frac{\hat\Sigma^{\gamma}}{s},\
\Pi^{Z}=\frac{\hat\Sigma^{Z}}{s-m_Z^2},\
\Pi^{\gamma Z}=\frac{\hat\Sigma^{\gamma Z}}{s}.$$

\begin{figure*}
\vspace*{20mm}
\hspace*{-5mm}
\scalebox{0.4}{\includegraphics{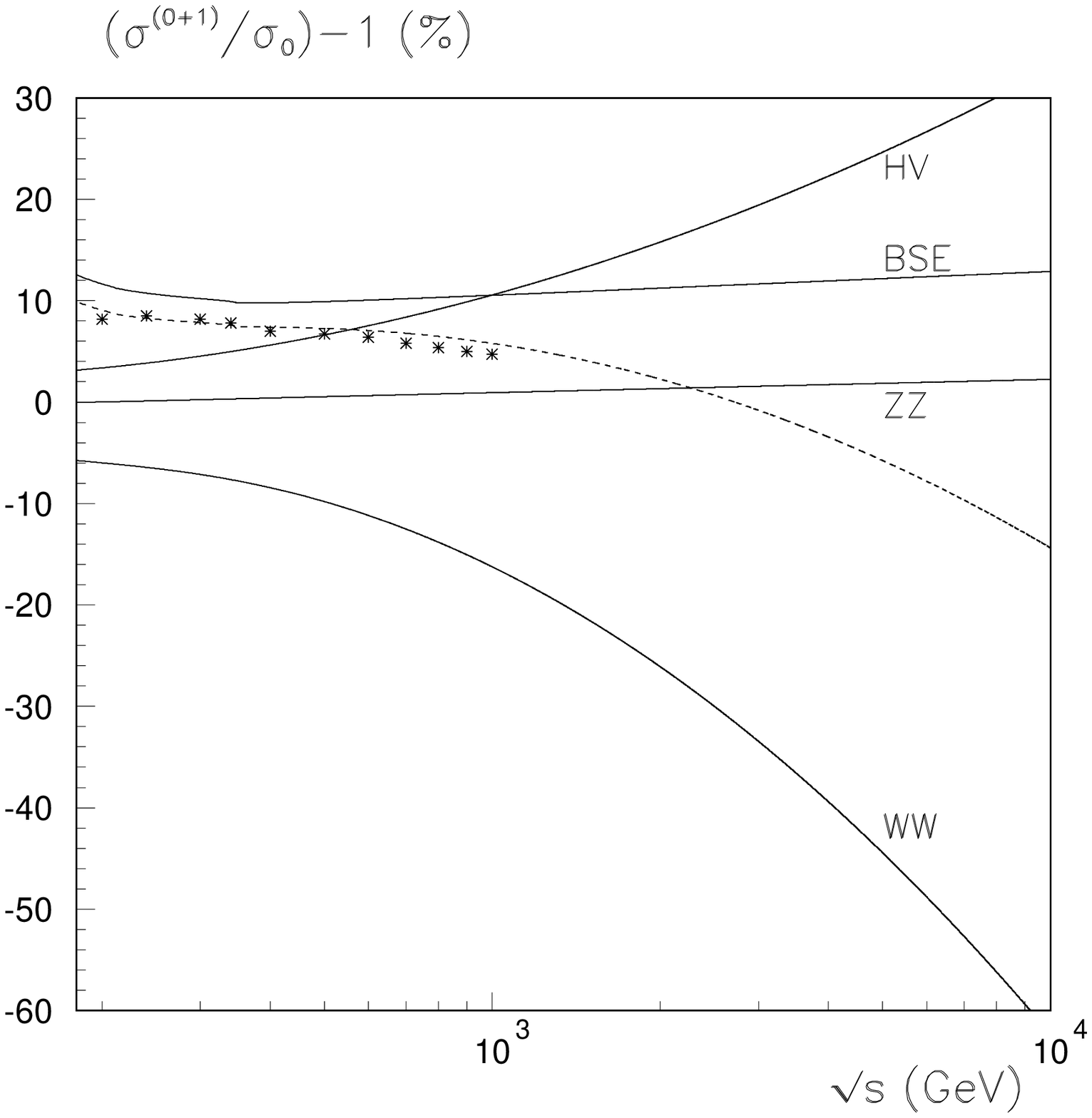}}
\scalebox{0.4}{\includegraphics{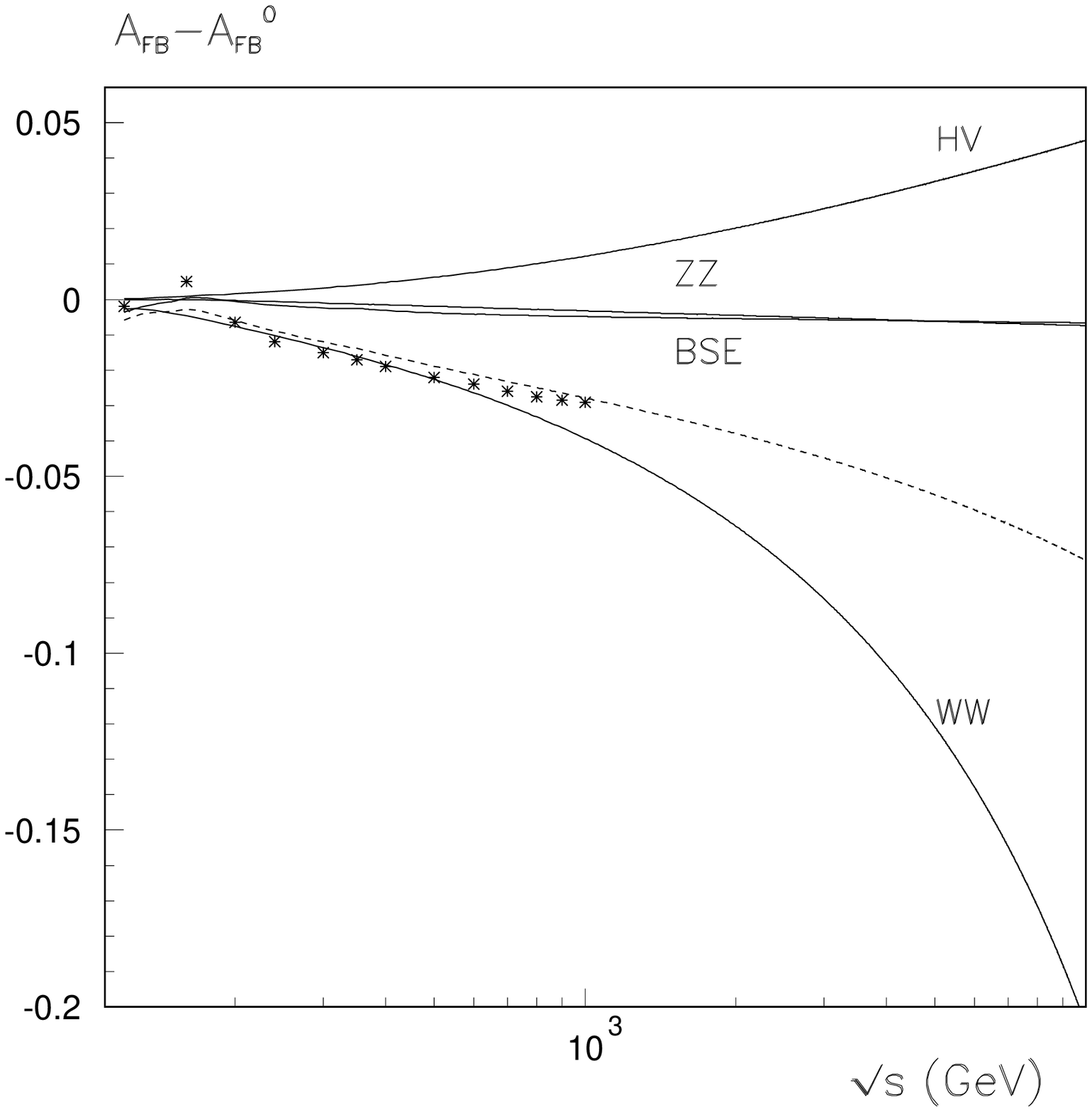}}
\vspace*{-5mm}
\caption{\label{fig:2}
The relative corrections to the total cross section (the left picture)
and the forward-backward asymmetry at the parton level (the right picture)
for $u\bar u \rightarrow \mu^+\mu^-$
as a functions of $\sqrt{s}$.
The symbols BSE, HV, ZZ, WW denote the
boson self energy, heavy vertex, ZZ-box, WW-box
contributions correspondingly.
The dashed line correspons to the sum of all contributions.
The asterisks are the points taken from FIG.~3 and FIG.~4 
of paper \cite{DY2002}.
}
\end{figure*}

Thus, Fig. 2 shows the relative corrections to the total
cross sections (the left picture) and forward-backward asymmetry
at the parton level (the right picture)
for $u\bar u \rightarrow \mu^+\mu^-$
as the functions of $\sqrt{s}$. As in \cite{DY2002}, for $e^+e^-$
final state we obtain the identical results.
The distinction between the $d\bar d$-- and
$u\bar u$--scattering is the same as in \cite{DY2002}, therefore
we do not present the graphs for $d\bar d$--case here.
The symbols BSE, HV, ZZ, WW in Fig. 2 denote the
boson self energy, heavy vertex, ZZ-box, WW-box
contributions correspondingly, the dashed line corresponds
to the sum of all contributions.
The asterisks are the points taken from FIG. 3 and 4 of paper \cite{DY2002}.

A rather good coincidence with the
presented results and the results  available from \cite{DY2002} takes place:
we can see in Fig. 2 almost the same scale and behavior
of relative corrections (the decrease with energy, the convexity
and t-quark threshold kink) and the tendency of the results to converge 
with the increase of energy.
We remind that the approaches in both calculations are not quite similar.
Although for the calculation of the BSE and HV parts
here and in Ref.\cite{DY2002} the same BSE insertions and
form factors have been used, but the heavy box contribution
has been calculated in different ways:
via the direct 4-point functions integration in \cite{DY2002},
and using asymptotic property of heavy boxes expressed
through the Sudakov logarithms powers in this paper.
It is obvious that the difference of the corrections 
in different approaches
strongly depends on the scale of energy: the asymptotic Sudakov expansion
works well at large values of $\sqrt{s}$ only.
Fig. 2 shows that at $\sqrt{s}$=1~TeV this difference  is rather small
both for total $u\bar u$ cross section ($\sim $ 1.5\%)
and for quark forward-backward asymmetry ($\sim $ 0.001)
and with the increase of $\sqrt{s}$ is found to become still smaller. 
It is a pity, but in \cite{DY2002} the parton observables 
in the region $\sqrt{s} > $~1~TeV have not been investigated.

To compare our results with the available 
in the region of high invariant masses $\sqrt{s} > $~1~TeV
we contacted to SANC group \cite{SANC}
and to the authors of \cite{DY2002} (see also program ZGRAD \cite{ZGRAD}).
Having used the same sets of electroweak parameters 
we got excellent coincidence for all WRC (see Table 1),
which shows the same (as in Fig. 2) 
corrections (in per cents) to $u \bar{u}$-cross section
as a functions of $\sqrt{s}$ in the region
$0.1~\mbox{TeV}~\leq~\sqrt{s}~\leq~10~\mbox{TeV}$.

\ \\
\ \\
\ \\
{
\begin{tabular}{|c||c|c|c||c|c||c|c|c||c|c|}
\hline
 \multicolumn{1}{|c||}{$\sqrt{s}$,\ TeV} 
& \multicolumn{1}{c|}{ZZ \cite{SANC}} & \multicolumn{1}{c|}{ZZ \cite{ZGRAD}}
 & \multicolumn{1}{|c||}{ZZ,\ AA} 
& \multicolumn{1}{c|}{WW \cite{SANC}} & \multicolumn{1}{c||}{WW,\ AA}
& \multicolumn{1}{c|}{HV \cite{SANC}}  
& \multicolumn{1}{c|}{HV,\ $\Lambda$} 
& \multicolumn{1}{c||}{HV,\ (\ref{L23})}
& \multicolumn{1}{c|}{BSE \cite{SANC}} & \multicolumn{1}{c|}{BSE,\ (\ref{bse})}\\
\hline 
 0.1 &   -0.0186 &           &   0.0683 & -0.329  &  -1.2690& -1.5549 &          &  2.2972  &   6.0777 &   8.1119 \\
 0.2 &   -0.0908 &   -0.0907 &  -0.0073 & -3.107  &  -4.8739& -1.6883 &  -1.6688 &  2.4481  &  11.2259 &  12.2144 \\
 0.5 &   -0.2144 &   -0.2145 &  -0.1895 & -10.777 & -10.1087&  4.2958 &   4.2978 &  4.0943  &  11.1526 &  11.9455 \\
 1.0 &   -0.3346 &   -0.3346 &  -0.3251 & -16.998 & -16.5720&  6.2447 &   6.2451 &  5.9910  &  12.2096 &  12.9793 \\
 2.0 &   -0.4638 &   -0.465  &  -0.4612 & -25.442 & -25.2497&  8.5534 &   8.5535 &  8.4247  &  13.1993 &  13.9634 \\
 3.0 &   -0.5423 &   -0.543  &  -0.5410 & -31.468 & -31.3554& 10.1709 &  10.1710 &  10.0935 &  13.7682 &  14.5314 \\
 5.0 &   -0.6432 &   -0.643  &  -0.6416 & -40.185 & -40.1306& 12.4894 &  12.4895 &  12.4512 &  14.4811 &  15.2437 \\
10.0 &   -0.7787 &   -0.779  &  -0.7782 & -53.989 & -53.9695& 16.1160 &  16.1160 &  16.1024 &  15.4456 &  16.2080 \\
\hline
\end{tabular}
}
\ \\
\ \\
\ \\
Table 1. The relative corrections (in percents) 
to the $\rm ZZ$, $\rm WW$, $\rm HV$ and $\rm BSE$ cross sections 
at the parton level for $u\bar u \rightarrow \mu^+\mu^-$
as functions of $\sqrt{s}$, calculated by different groups:
SANC\cite{SANC}, program ZGRAD \cite{ZGRAD}
and using the results of this paper. 
\ \\
\ \\
\ \\

Let us analyse the ZZ-box:
we can see that in the region of not so large values of
 $\sqrt{s}$ (0.1 -- 0.2~TeV)
the results are rather different (although the corrections 
hold the similar behaviour and scale), but starting with
 $\sqrt{s} \sim 0.5$~TeV the difference is small and tends to zero
with  the growth of $\sqrt{s}$. Of course, this fact confirms
the precision of both calculations and estimates the 
region where the asymptotic approach suggested here works well.
I suppose this point is $0.5$~TeV, here the relative error
of asymptotic (to exact SANC) calculation is $\sim 0.1161$, whereas
at $1$~TeV ($2$~TeV) the error is $\sim 0.0284$ ($\sim 0.0056$).  

There is no special need to compare separately the direct and crossed
parts of ZZ-box correction (the same way as WW-box)
containing DSL,
since having the good agreement for the whole ZZ-box,
which is proportional merely to the first power of Sudakov logarithms 
(see (\ref{diff})),
we will obtain the agreement of the same order.
Nevertheless, we compare the results for the direct WW-box:
for the point $0.5$~TeV the relative error
of asymptotic (to SANC) calculation is $\sim 0.0620$, and
at $1$~TeV ($2$~TeV) this error is $\sim 0.0251$ ($\sim 0.0076$).  
Table 1 also shows good agreement for HV part of radiative correction: 
the first HV column is the result of SANC, the second one is obtained
using the formulas of Section III and exact expessions for 
$\Lambda_{2,3}$ from \cite{BSH86}, the third column corresponds
to the approximation (\ref{L23}); we can see a very good coincidence
of the first and second columns and a rather good agreement between
the first and third columns -- the relative error is 0.040 at $\sqrt{s}$=1~TeV 
and $8.44 \times 10^{-4}$ at $\sqrt{s}$=10~TeV.
For the BSE part the agreement is not so good as for HV --
there is a small constant difference $\sim 0.5\%$ obviously 
caused by the use of various gauges.

Let us pass to the numerical analysis of hadron observables.

\subsection{Numerical estimation for future CMS program}
Futher the scale of weak radiative corrections and their effect
on the observables of the Drell-Yan processes for future CMS experiments
will be discussed.
We used the following set of electroweak parameters (such as in \cite{ZGRAD}):
$\alpha=1/137.03599911$,\ 
$m_W=80.37399\ \mbox{GeV}$,\ $m_Z=91.1876\ \mbox{GeV}$,
the energy of LHC $\sqrt{S}=14\ \mbox{TeV}$
and the CTEQ6 set of unpolarized parton distribution functions (PDF)
\cite{CTEQ6} (with the choice $Q^2=M^2$).

We impose the experimental restriction conditions
on the detected lepton angle
$-\zeta^* \leq \zeta \leq \zeta^*$
and on the rapidity 
$|y(l)|\leq y(l)^*$.
For CMS detector the cut values of $\zeta^*$ and $y(l)^*$
are determined as
\begin{equation}
y(l)^* = - \ln \ \tan \frac{\theta^*}{2} = 2.4,\
\zeta^* = \cos\theta^* \approx 0.9837.
\label{restr}
\end{equation}
Then the expression (\ref{xsM}) with consideration of
the experimental restrictions (\ref{restr}) 
and the use of theta-functions
can be modified to
\begin{equation}
\sigma(M)=\int\limits_{-\ln\frac{\sqrt{S}}{M}}^{+\ln\frac{\sqrt{S}}{M}} dy
          \int\limits_{-\zeta^*}^{\zeta^*} d\zeta\
          \sigma(M,y,\zeta) \theta(\cos\alpha+\zeta^*)
          \theta(-\cos\alpha+\zeta^*),
\label{xsMr}
\end{equation}
where $\alpha$ is the scattering angle of the lepton
with the 4-momenta $k_2$
($\alpha = \widehat{\vec p_1, \vec k_2}$) in the center
mass system of hadrons.
This angle has a non-trivial relation with $\theta$ and $y$:
\begin{equation}
\alpha = \pi - \arccos \frac{\cos\theta - f}{\sqrt{1+f^2-2f\cos\theta}}
 - \arcsin \frac{f \sin\theta}{\sqrt{1+f^2-2f\cos\theta}},\
f=\frac{e^{2y}-1}{e^{2y}+1}.
\end{equation}
Let us note here that the second standard CMS restriction
$p_T(l) \geq 20\ \mbox{GeV}$ in the region of large $M$
considered here
and at (\ref{restr}) is satisfied completely and automatically.
It can be seen in Fig.3
where all of the CMS experimental restrictions for $M=1\ \mbox{TeV}$ 
have been  shown.

\begin{figure*}
\vspace*{15mm}
\hspace*{-5mm}
\scalebox{0.4}{\includegraphics{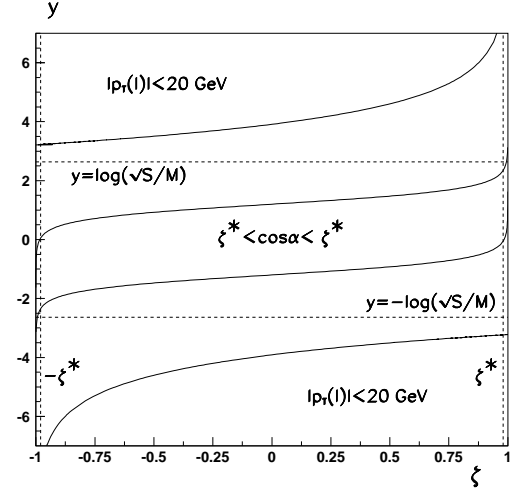}}
\vspace*{-15mm}
\caption{\label{fig:3}
The region of the cross section integration over $\zeta$ and $y$
taking into consideration the CMS experimental cuts
at $\sqrt{S}=14$~TeV and $M$=1~TeV.
}
\end{figure*}

\begin{figure*}
\vspace*{45mm}
\hspace*{-15mm}
\scalebox{0.4}{\includegraphics{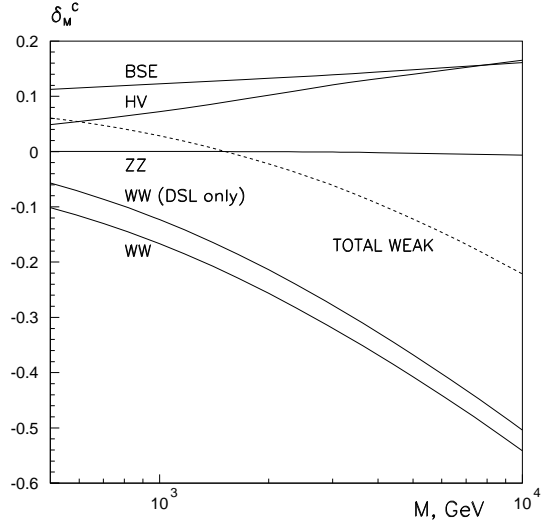}}
\vspace*{-15mm}
\caption{\label{fig:4}
The relative corrections $\delta_{M}^{C}$
corresponding to
the $\rm ZZ, WW, HV, BSE$--contributions
with experimental restrictions of the CMS
as a functions of $M$.
The dashed line corresponds to the sum of all the WRC contributions.
}
\end{figure*}

In Fig.4  the relative corrections are presented
\begin{eqnarray}
\delta^{C}_{M} = \frac{\sigma_{C}(M)}{\sigma_0(M)},
\label{de}
\end{eqnarray}
as  functions of $M$ in the region
$0.5~\mbox{TeV}~\leq~M~\leq~10~\mbox{TeV}$
with experimental restrictions of the CMS.
Index "$C$" in all numerical analysis means
\begin{eqnarray}
C={\rm ZZ, WW, HV, BSE, tot;\ \ tot=ZZ+WW+HV+BSE}.
\end{eqnarray}
It can be seen, that the BSE, HV, and WW--contributions are most significant,
and the WW boxes give the dominant (negative) contribution.
In Fig.4 we can also see the significant contribution of zero power
of Sudakov logs ($\sim l^0_{W,x}$): the difference 
"WW (DSL only)" - "WW (total)" is $\sim 4\%$ and depends weakly on $M$.
Big contribution of $l^0_{W,x}$  justifies
the detailed calculations which have been done in Section IV.
The $4\%$--difference is caused mainly by the $2\pi^2/3$--term 
in (\ref{d1}), the contribution of $D_2^{WW}$ (\ref{d2}) is far less than 1\%.
The total WRC effect is significant: in region $M < 1.5~\mbox{TeV}$
is positive, since the sum of positive HV and BSE contributions
exceeds the negative WW-boxes, but for $M > 2~\mbox{TeV}$ 
the total correction is definitely negative and decreases
with the increase of $M$. 
Thus, the $\delta_{M}^{tot}$ correction
changes the dilepton mass distribution
up to $\sim +3 (-12) \% $ at $M=1(5)~\mbox{TeV}$.

Here we should note that the importance  of the investigated problem
demands the cross-checking of this part with the earlier obtained 
in \cite{DY2002}.
The direct comparison between numerical results of the two works
is not possible, since in  Ref.\cite{DY2002} 
the effective Born approximation has been used to calculate 
the relative corrections.
We do not use the effective Born approximation as
the main subject of our investigation is the region
of large invariant mass ($M \geq $~1~TeV)
where this approach is illegal. 
Fortunatly, recent paper \cite{Baur2006} provides the possibility to compare 
the results of Ref.\cite{DY2002} and presented by us in the region
of high invariant dilepton mass (see the first picture 
in Fig.7 of \cite{Baur2006}). At first glance the results look
absolutely different: for $M=1(2)$~TeV the relative corrections
of \cite{Baur2006} is $\sim -5\% (-11\%)$ whereas our result 
is $\sim +3\% (-2\%)$.
However, the solid line in Fig.7 of \cite{Baur2006} correspons to total
one-loop result and to compare correctly we should add  
the photon corrections: "soft" and "hard" photon contributions.
We have our own code for calculation of this part (FORTRAN program
READY -- "Radiative corrEctions to lArge invariant mass Drell-Yan process",
which is supposed to be published in the near future, for more details 
see \cite{arx07}). 
Main features of READY are the same as in Ref.\cite{DY2002}:
using the "soft"-"hard" separator $\omega$ and independence of
total result on it, lepton identification cuts 
reduce extremely the "hard" FSR contribution and so on.
For the decision of quark mass singularity problem we used
the procedure of linearization \cite{SANC}.
The "soft" part, which we should add to pure WRC for comparison with
\cite{Baur2006}, is large and negative (for example,
at $M=1$~TeV, $\omega=10$~GeV and $l=e$ the FSR-part of
"soft" relative correction defined by formulas (72),(73) 
in Ref.\cite{YAFDY} equals $-40.7\%$. The positive "hard"
FSR-part suppressed by lepton identification requirements
will be far less, and the rest parts (ISR after quark mass removing,
and the mass-independing interference) will be small as 
compared to the FSR. 
Hence, we are supposed to have a rather good 
(at least qualitative) agreement
with hadron estimations of Ref.\cite{Baur2006}: the relative correction
will be negative and decrease with $M$ in region $M \gtrsim $~0.5~TeV.
Unfortunatly the READY is not quite complete and here we can not present 
more exact numbers, but we hope future publications of the results of 
code READY must clarify this problem later.

In the end of this section we want to discuss
the effect of the PDFs on the weak radiative 
corrections in this very high mass
regime, and what uncertainty they introduce. 
Comparing the results for CTEQ6 set \cite{CTEQ6}
and MRST 2004-QED set \cite{MRST} of PDFs
the essential effect of PDF choice for the cross sections
has not been noticed, the relative error is on one per cent level
and has no tendency to increase with the increase
of $M$, at least at $M \sim 3.16$~TeV
(unfortunately MRST 2004-QED set for $Q^2 > 10^7$ GeV,
i.e. for $M > 3.16$~TeV, does not work, and 
it does not allow to compare the result for both sets of PDF
in the region of extremely high invariant masses).

 \section{Conclusions}

The weak radiative corrections with the neutral current
to the Drell-Yan process
for large invariant mass of a dilepton pair have been studied.
The compact asymptotic expressions have been obtained, which
expand in the powers of the Sudakov electroweak logarithms.
At the parton level the investigated
radiative corrections have been compared with the existing results
and a rather good coincidence at energy $\gtrsim $~0.5~TeV
have been found.
The numerical analysis in the high energy region has been performed
with use of the standard CMS cuts.
It has been found that the considered
radiative corrections become very significant at high dilepton mass,
and their large scale does not permit to neglect this part of
the radiative correction procedure
in the future experiments at LHC directed on the NP research
in the Drell-Yan process.
Some part of corrections has become beyond the scope
of the presented paper (the QCD corrections, the pure QED corrections,
and the two-loop electroweak logarithms), and still
it requires a thorough analysis in order for us to completely solve
the radiative corrections problem
to the Drell-Yan process at extremely high energies.

\section{Acknowledgments}
I would like to thank A.~Arbuzov, D.~Bardin, I.~Golutvin, E.~Kuraev, 
V.~Mossolov, S.~Shmatov, N.~Shumeiko for the stimulating discussions.
I am grateful to A.~Arbuzov, D.~Bardin, S.~Bondarenko and D.~Wackeroth for
a detailed comparison of the results.
I thank CERN (CMS Group), where part of this work has been carried out, 
for warm hospitality during my visits.

\end{document}